\documentclass[12pt,a4paper]{article} 
\usepackage{graphicx} 
\usepackage{setspace}
\usepackage[paper=a4paper,left=30mm,right=30mm,top=30mm,bottom=30mm]{geometry}
\usepackage[footnotesize]{caption}
\begin{document}
\renewcommand{\thefootnote}{\fnsymbol{footnote}}
\thispagestyle{empty}
\vspace*{1cm}
\begin{center}
 \section*{\bf Controlling tax evasion fluctuations\footnote[1]{We thank Dietrich Stauffer for his stimulating discussions and many constructive comments.}}
\vspace{0.5cm}

Frank Westerhoff\footnote[4]{Department of Economics, University of
  Bamberg, Feldkirchenstrasse 21, D-96045 Bamberg, Germany.} \\
F.W.S. Lima\footnote[3]{Departamento de F\'{i}sica, Universidade Federal do Piau\'{i}, 64049-550 Teresina -- PI, Brazil.}\\
Georg Zaklan$^{\S}$\\[0.3cm]
\vspace{0.5cm}
e-mail:   georg.zaklan@uni-bamberg.de\\[1cm]   
\vspace{0.5cm}
\today 
\end{center}

\vspace{0.55cm} 
  \begin{abstract}

\noindent We incorporate the behaviour of tax evasion into the standard two-dimensional Ising model and augment it by providing policy-makers with the opportunity to curb tax evasion via an appropriate enforcement mechanism. We discuss different network structures in which tax evasion may vary greatly over time if no measures of control are taken. Furthermore, we show that even minimal enforcement levels may help to alleviate this problem substantially. 
\\[0.3cm]
\noindent [Keywords: Opinion dynamics, 
Sociophysics, Ising model.]\\[0.0cm]
\end{abstract}
\newpage
\renewcommand{\thefootnote}{\arabic{footnote}}
\section{Introduction}
In economics, the problem of tax evasion from a multi-agent-based perspective has received little attention so far (see Bloomquist (2006) for a recent overview). Realistic models on tax evasion appear to be necessary, because tax evasion remains to be a major predicament facing governments (see Andreoni et al., 1998; Lederman, 2003; Slemrod, 2007).

Experimental evidence (see G\"achter, 2006) suggests that tax payers usually condition their decision regarding whether to pay taxes or not on the tax evasion decision of the members of their group (``conditional cooperation''). Conditional cooperators are more likely to evade taxes if they have the impression that many others evade. On the other hand if most others behave honestly, an individual is less likely to cheat on her taxes. G\"achter  presents the findings of public goods experiments  and argues that conditional cooperation primarily motivates people to either contribute to the provision of a public good or to free-ride. 
Frey and Torgler (2006) provide empirical evidence on the relevance of conditional cooperation for tax morale. They find a positive correlation between people's tax morale, which is measured by asking whether tax evasion is justified if the chance arises, and their perception regarding how many others evade paying tax.
Conditional cooperation from the viewpoint of the standard economic theory may be explained by changes in risk aversion due to changes in equity (Falkinger, 1995).

We decide to use the Ising model, because it allows to model conditional cooperation in a multi-agent-based fashion. It allows to consider a large number of agents who interact locally with each other and base their decision whether to evade taxes or not on the behaviour of the other agents in their group.

We incorporate the behaviour of tax evasion and add an enforcement mechanism into the standard two-dimensional square lattice Ising spin model. We aim to extend the study of Zaklan et al. (2008), which illustrates how, in a world where agents are conditionally cooperative, different levels of enforcement affect aggregate tax evasion over time. We define enforcement to consist of two components: a probability of an audit ($p$) each person is subject to in every period and a length of time detected tax evaders need to remain honest for ($k$ periods). We embed our tax evasion model into different network structures and find for these networks that fluctuation in aggregate tax evasion behaviour may arise if no enforcement is used. 
For our simulations we make use of the ``tunnelling'' process at temperatures slightly below the critical temperature.
This process has been used for two decades in the field of physics, and also by Hohnisch et al. (2005) for the IFO-Index. We use it to illustrate a second important and maybe less obvious effect of enforcement, as we define it: we provide evidence, by simulation, that even minimal levels of enforcement may help to reduce the presence of fluctuations in tax evasion. Such fluctuations can be completely prevented in the considered networks by setting the enforcement measures to sufficiently high, but realistic, levels. Everybody then remains compliant for most of the time.

The remainder of our manuscript is organised as follows: in section 2 we present our model, which is based on the standard two-dimensional Ising model on a square lattice. In section 3 we decribe the evolution of the aggregate tax evasion behaviour that our model generates under different enforcement regimes. In section 4 we additionally embed our model into the Barab\'asi-Albert network and the Voronoi-Delaunay random lattice and 
discuss the resulting tax evasion dynamics.

\section{The model}
We use the standard Glauber kinetics of the Ising model on a 20 $\times$ 20 square lattice (in section 4 we will analyse our model for other lattice types). In every time period each lattice site is occupied by an individual who can either be an honest tax payer ($S_i = +1$) or a tax evader ($S_i =-1$). The small number of agents may be imagined to represent the elite of a country, whose tax evasion behaviour it may be interesting to look at, given the different enforcement regimes of the tax authority. For our analysis we assume that  everybody is honest initially. Each period individuals have the opportunity to become the opposite type of agent as they were in the previous period. Each agent's social network, which is made up of four next neighbours, may either prefer tax evasion, reject it or be indifferent.

Tax evaders have the greatest influence to turn honest citizens into tax
evaders if they constitute a majority in
 the given neighbourhood. 
If the majority evades, one is likely to 
also evade. On the other
hand, if most people in the vicinity are 
honest, the respective
individual is likely to become a decent citizen
 if she was a tax evader
before. How strong the influence from 
the neighbourhood is can be
controlled by adjusting the temperature, $T$.
Total energy is given by the Hamiltonian 
$H=-\sum_{<i,j>}J_{ij} S_i S_j-B\sum_{i}S_i$. We choose $J=1$ and $B=0$. 
For very low temperatures, the
autonomous part of decision-making almost
completely disappears.\footnote{The autonomous
part of individual decision-making 
is responsible for the emergence of
the tax evasion problem, because some
 initially honest tax payers
decide to evade taxes and then exert
 influence on others to do so as
well.} 
Individuals then base their 
decisions solely on what most
of their neighbours do. A rising temperature 
has the opposite
effect. Individuals then decide more
autonomously. It is well known that for
$T > T_c$ ($\approx 2.269$), half of the people are honest and 
the other half cheat, while for $T < T_c$ states coordinated on cheating or compliance prevail for most of the time.

As an enforcement measure, we introduce a probability of an efficient audit ($p$). If tax evasion is detected, the individual must remain honest for a certain number of periods. We denote the period of time for which detected tax evaders are punished by the variable $k$. One time unit is one sweep through the entire lattice. Audits are stochastically independent from other agents and from the history any agent has.

\section{Dynamics of the model}

\vspace{0.3cm}
\begin{center}
---$\quad$Figure $1$ goes about here$\quad$---
\end{center}
\vspace{0.3cm}
The top-left panel of Figure 1 illustrates the baseline setting, i.e. no use of enforcement, for the square lattice.  We depict the dynamics of tax evasion over $50,000$ time steps.
Although everybody is honest initially, it is not possible to predict which level of tax compliance will be reached at some time step in the future. Agents are usually either mostly compliant or mostly non-compliant, whereas the system typically remains in either state for a while. Switching from a mostly compliant to a mostly non-compliant society, or vice versa, is favoured by both the small number of agents and the temperature, which needs to be somewhere close to the critical level (we use $T\approx T_c$). If, by chance, more than 50\% of agents start to prefer the opposite action of the currently dominating one, this strategy will then start to prevail for a while. As soon as there is a majority for the previously dominating strategy regarding tax evasion, aggregate behaviour is then likely to reverse again. 
If more agents or a temperature further below the critical level are picked, it would take longer for a switch in aggregate evasion behaviour to occur. Apparently, a suitable measure of control is needed to prevent agents from repeatedly falling into non-compliance.
\vspace{0.3cm}
\begin{center}
---$\quad$Figure $2$ goes about here$\quad$---
\end{center}
\vspace{0.3cm}
Figure 2 illustrates different simulation settings for the square lattice, where for each considered combination of degree of punishment ($k=1$, $10$ and $50$)  and audit rate ($p=0.5$, $10$ and $90\%$)  the corresponding dynamics of tax evasion is depicted over $50,000$ time steps. Surprisingly, even very small levels of enforcement (e.g. $p=0.5\%$ and $k=1$) suffice to almost completely prevent fluctuation in aggregate tax evasion behaviour and to establish mainly compliance. 
Only seldomly tax evasion then becomes the predominant aggregate choice of action.
Both, a rise in audit probability (greater $p$) and a higher penalty (greater $k$), work to  flatten the time series of tax evasion and to shift the band of possible non-compliance values towards more compliance.
If the audit rate is increased to the level of 1\%, even for very small penalties ($k=1$) then an upsurge in tax evasion will not occur any longer (not displayed).
Since high income earners are audited more often ($p\approx 10\%$) than average income tax payers ($p\approx 1\%$), we look at how results change if higher levels of enforcement are used ($p=10\%$). 
Interestingly, higher audit rates only reduce the level of tax evasion marginally. The simulations illustrate that even extreme enforcement measures (e.g. $p=90\%$ and $k=50$) cannot fully resolve the problem of tax evasion.

\section{Modifications}
To examine whether the results generated by the square lattice are robust, we extend our analysis 
to other frequently used network structures. Specifically, we make use of the Voronoi-Delaunay random lattice and the Barab\'asi-Albert network model. The construction of the Voronoi-Delaunay lattice (i.e. tessellation of the plane for a given set of points) 
is defined as follows (Lima et al., 2000). For each point, one first needs to determine the polygonal cell, consisting of the region of space nearer to that point than to any other point. Whenever two such cells share an edge, they are considered to be neighbours.
From the Voronoi tessellation, one can obtain the dual lattice by the following procedure: when two cells are neighbours, a link is placed between the two points located in the cells. From the links, one obtains the triangulation of space. The network constructed in this manner, which we use for simulation, is called the Voronoi-Delaunay lattice. 

The Barab\'asi-Albert network (Barab\'asi and Albert, 1999) is grown such that the probability of a new site to be connected to one of the already existing sites is proportional to the number of connections the existing site has already accumulated over time: individuals with many friends are more likely to gain new friends than loners.
 

In these variations of our simple square lattice model we also choose 400 agents and  depict the resulting tax evasion dynamics over $50,000$ time steps.

The remaining pictures in Figure 1 illustrate the dynamics in the baseline setting in these additional network structures of our model. Both networks, the Voronoi-Delaunay lattice and the Barab\'asi-Albert network, support our findings in the case of the square lattice, namely that fluctuations in tax evasion behaviour may occur if no enforcement mechanism is implemented.

\vspace{0.3cm}
\begin{center}
---$\quad$Figure $3$ goes about here$\quad$---
\end{center}
\vspace{0.2cm}

\noindent Figure 3 illustrates the tax evasion dynamics for the Voroi-Delaunay lattice (first column) and the Barab\'asi-Albert (other two columns) network, if different degrees of enforcement are used.

For the Voronoi-Delaunay random lattice we also find that fluctuations in tax evasion can be reduced substantially by implementing very low probabilities of an audit. For an audit rate of $p=1\%$ no fluctuations occur any longer and society remains mainly honest over time.
Obviously higher punishments, i.e. higher levels of $k$, also lower the amount of non-compliance.

The next two columns illustrate that enforcement in the Barab\'asi-Albert network is less efficient than in the Voronoi-Delaunay network or in the 
simple square lattice. 
As can be seen, fluctuations still occur for $p=1\%$, even for high levels of punishment (e.g. $k=10$).
The last column illustrates the tax evasion dynamics, holding the audit rate constant at $p=4.5\%$. This is the minimal audit rate which (almost always) prevents tax evasion from fluctuating, even for the lowest considered level of punishment (i.e. $k=1$), and is much higher than in the other two network models, yet still at a realistic level.

\section{Conclusion}
Tax evasion can vary widely across nations, reaching extremely high values in some developing countries. Wintrobe and G\"erxhani (2004) explain the observed higher level of tax evasion in generally less developed countries with a lesser amount of trust that people accord to governmental institutions. Empirical evidence for the importance of trust for tax compliance may, for example, be found in Hyun (2005) and in Torgler (2004). So far we have neglected 
the effects of public opinion, which may vary greatly across countries, on tax evasion. It therefore seems worthwhile to extend the setting of our simple model in a further study, to specifically analyse the aspect of varying tax compliance across countries.

\newpage


\newpage
\setstretch {1.5}
 
\newpage
\pagestyle{empty}

\newpage
\vspace*{5.0cm}
\begin{center} 
\begin{tabular}{c}
\hspace{-0.8cm}
\hspace{-1.7cm}\includegraphics[width=6.8cm]{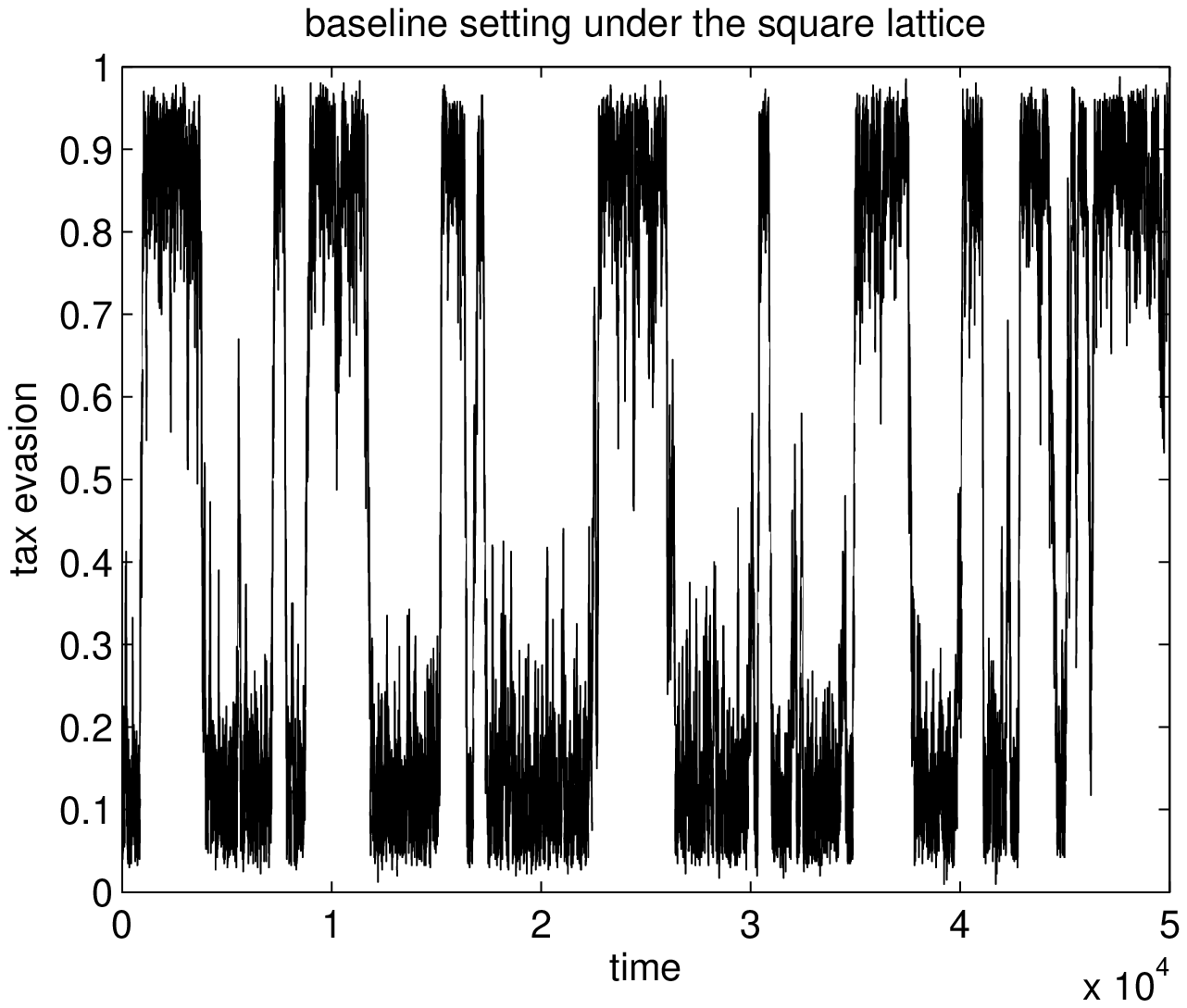}
\hspace{-0.75cm}
\includegraphics[width=6.8cm]{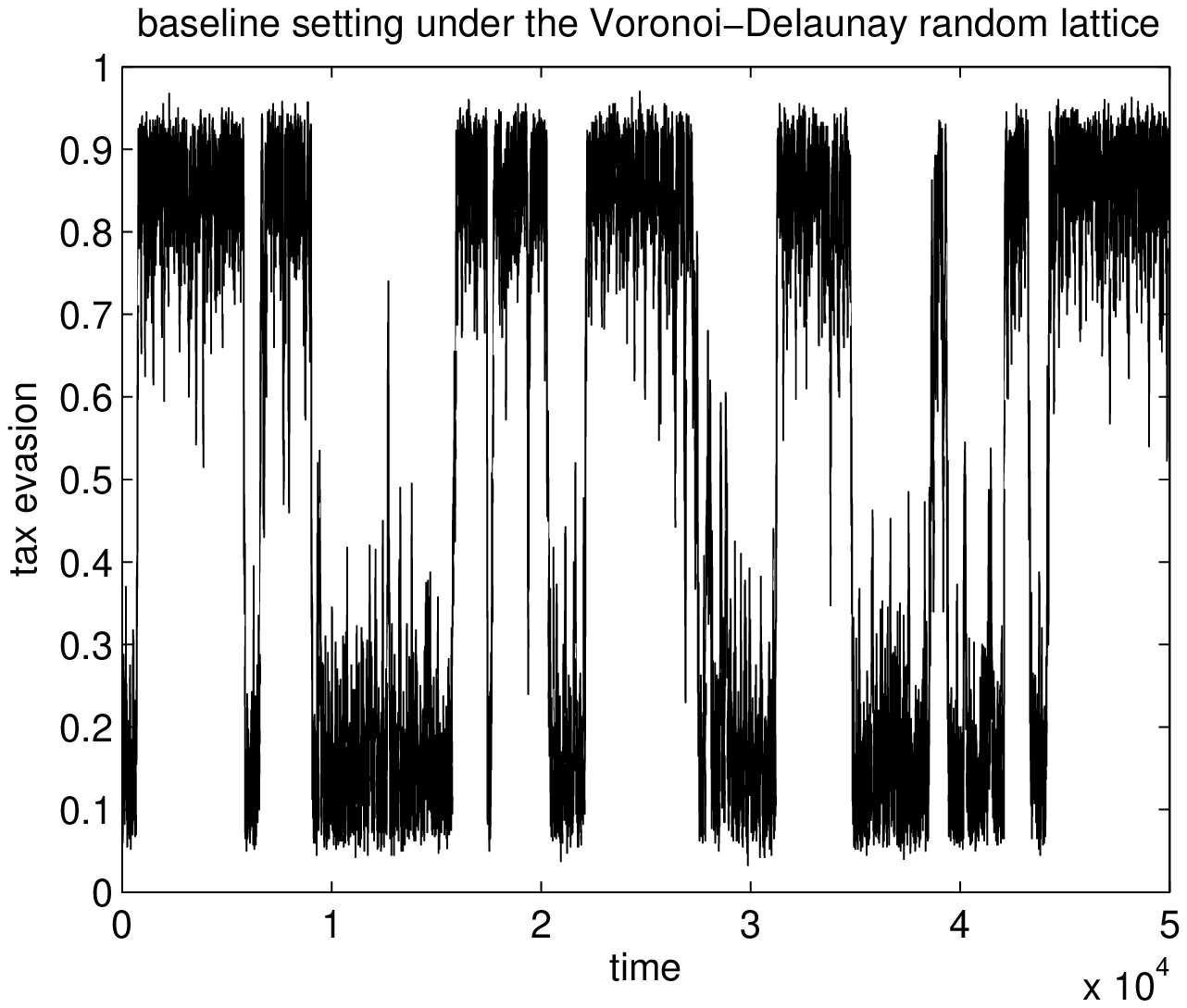}  
\hspace{-0.75cm}
\includegraphics[width=6.8cm]{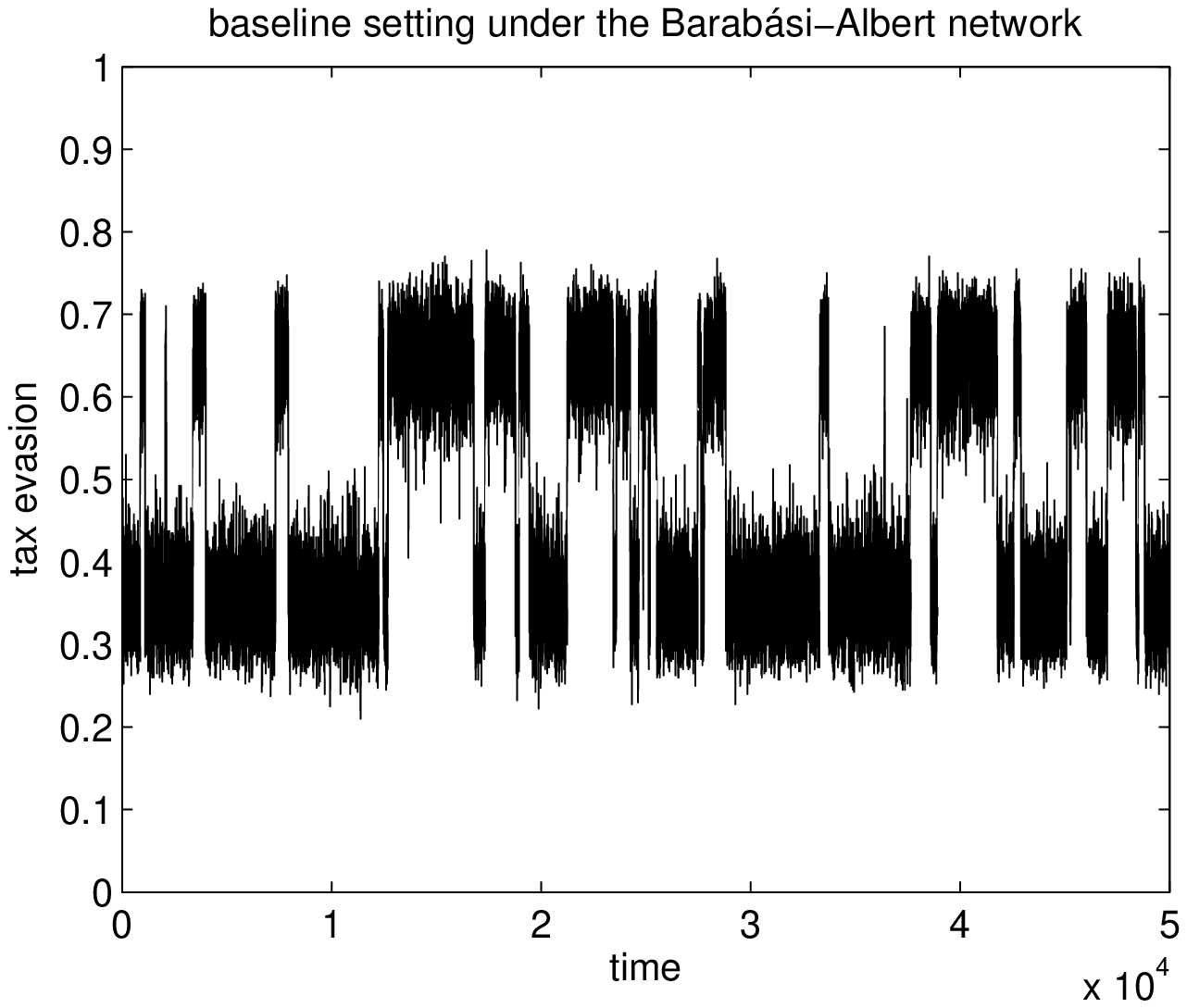} 
\end{tabular}
\end{center}
\vspace{0.3cm}
{
{\noindent Figure 1: The baseline setting is given if no enforcement measure is implemented to control tax evasion. The figures above illustrate the baseline settings for the different network structures we use. For the square lattice we take $T=2.265$, for the Voronoi-Delaunay lattice $T_c=3.802$ and for the Barab\'asi-Albert network $0.8\cdot T_c= 0.8 \cdot m\cdot \log(NSITES)/2$ (where $m=4$ and $NSITES=400$). All simulations are performed over $50,000$ time steps.}}

\newpage
\vspace*{1.2cm}
\begin{center} 
\begin{tabular}{c}
\hspace{-0.8cm}
\hspace{-1.7cm}\includegraphics[width=6.8cm]{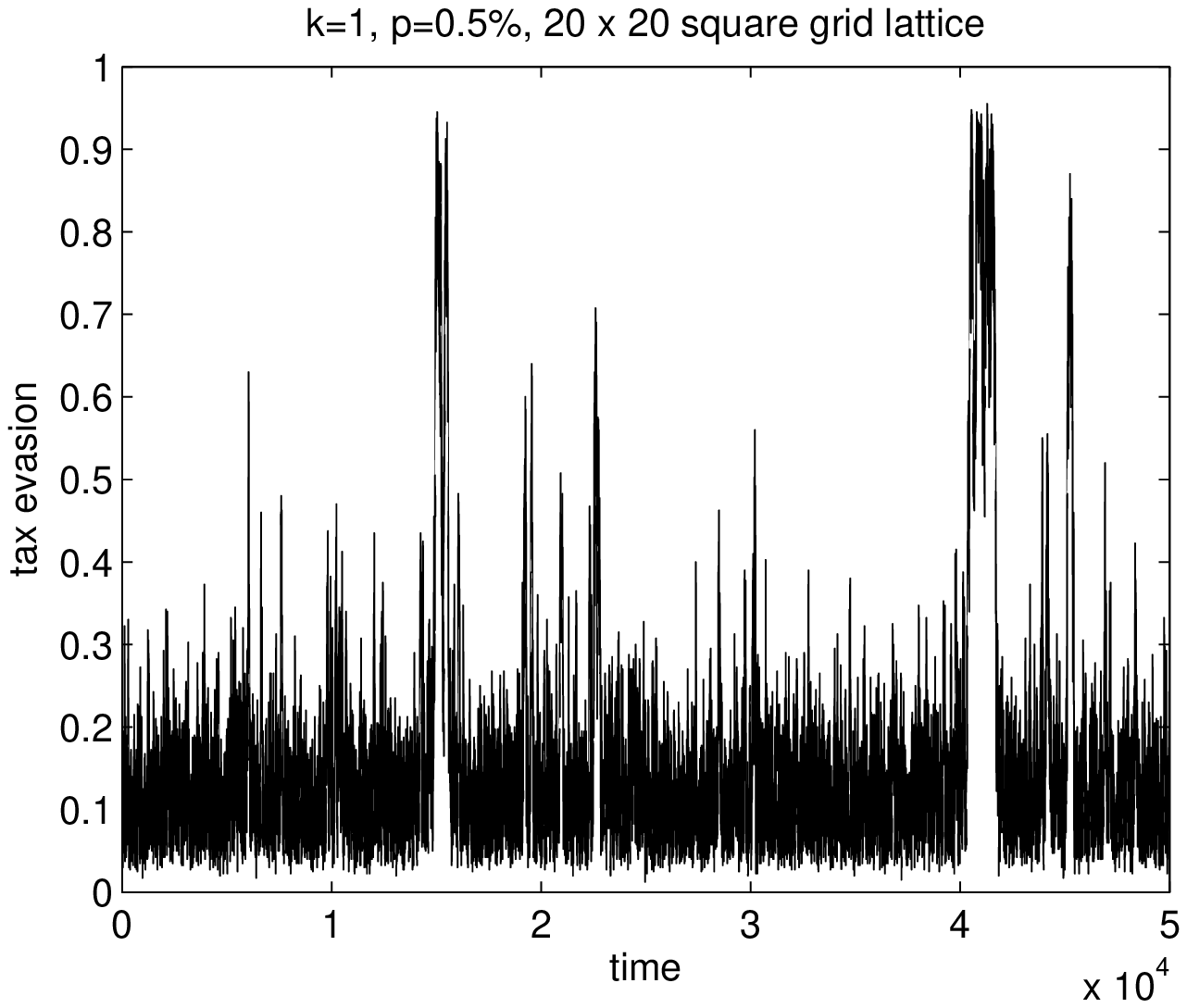}
\hspace{-0.75cm}
\includegraphics[width=6.8cm]{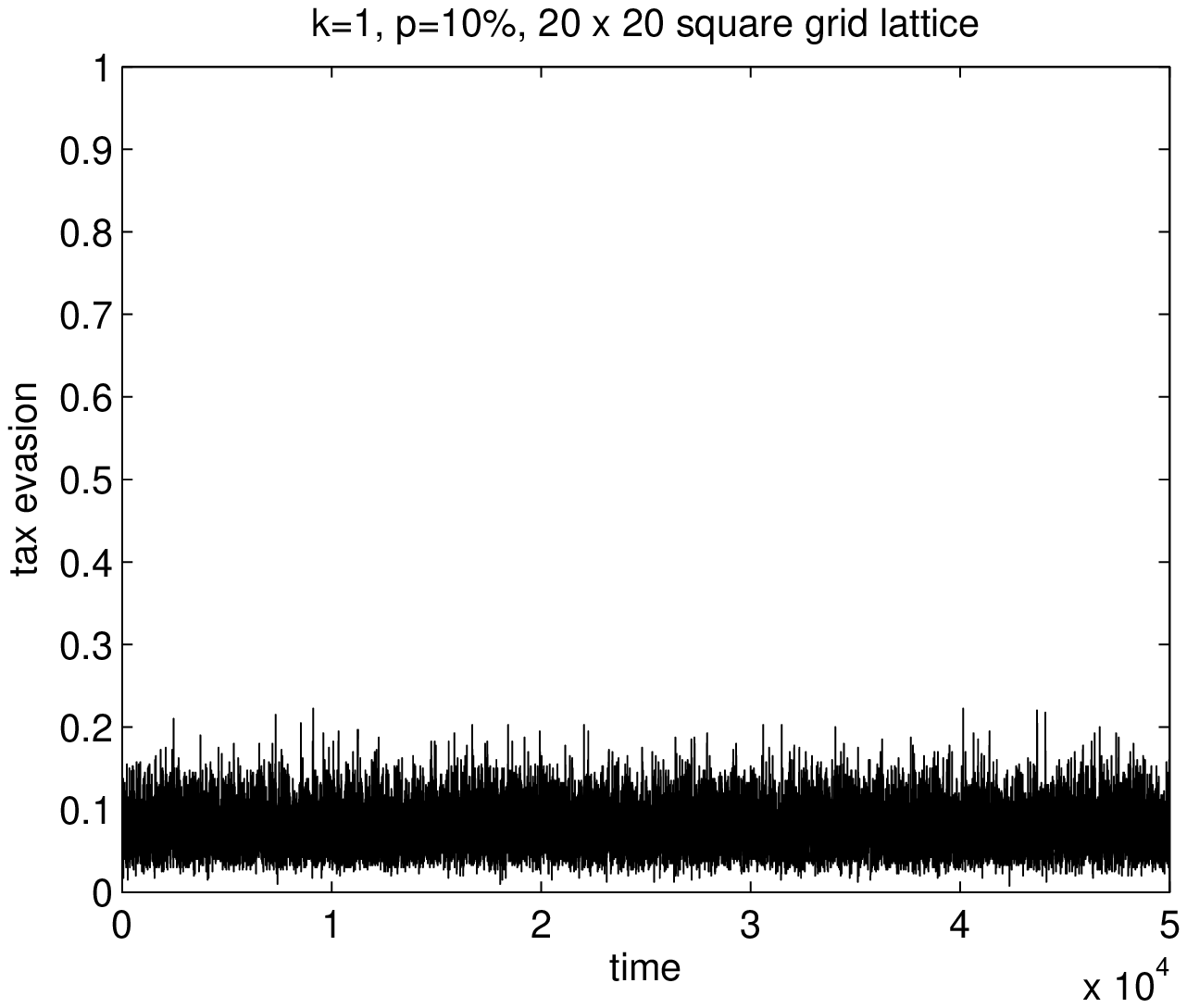}  
\hspace{-0.75cm}
\includegraphics[width=6.8cm]{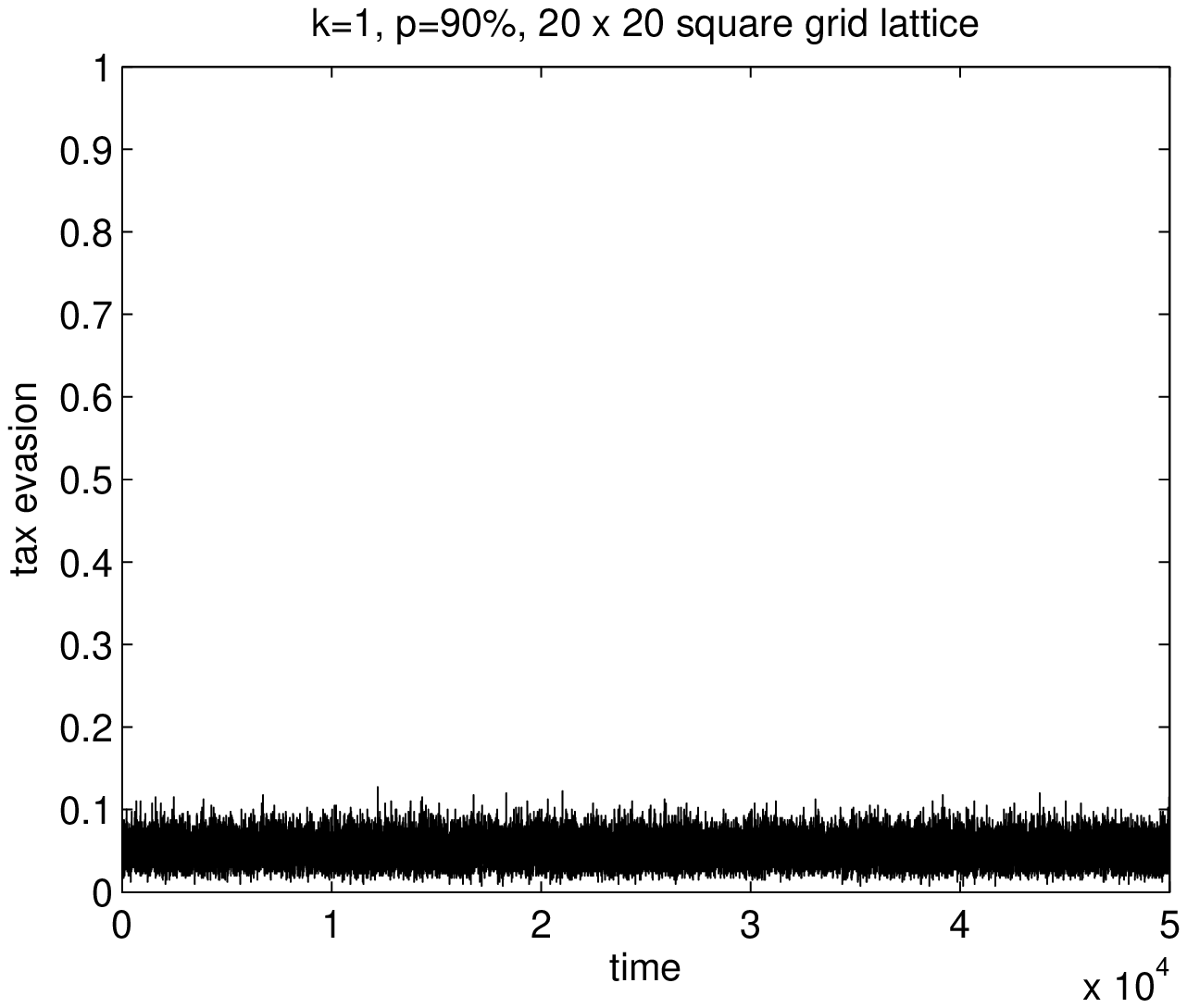} \\
\hspace{-0.8cm}
\hspace{-1.7cm}\includegraphics[width=6.8cm]{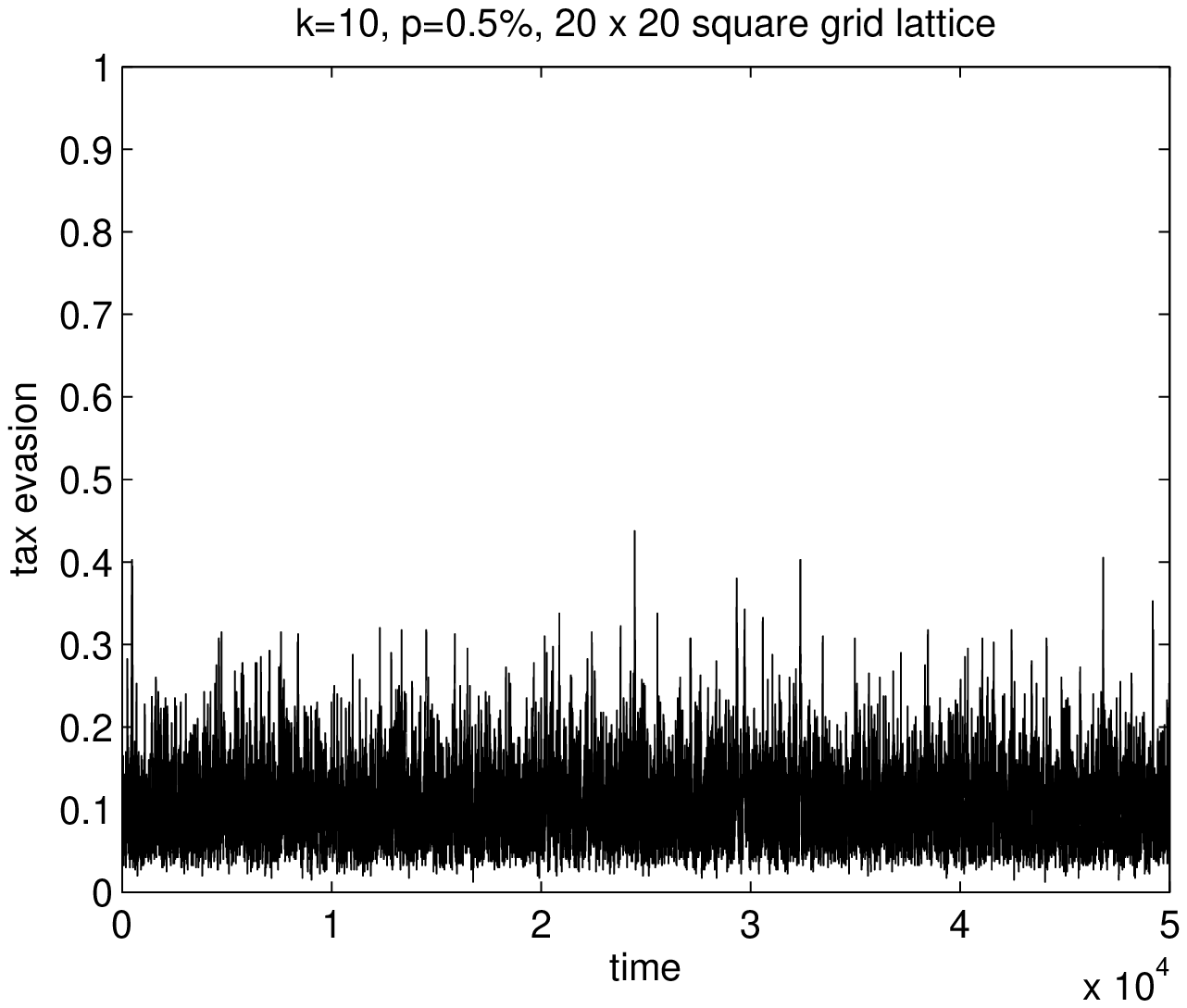}
\hspace{-0.75cm}
\includegraphics[width=6.8cm]{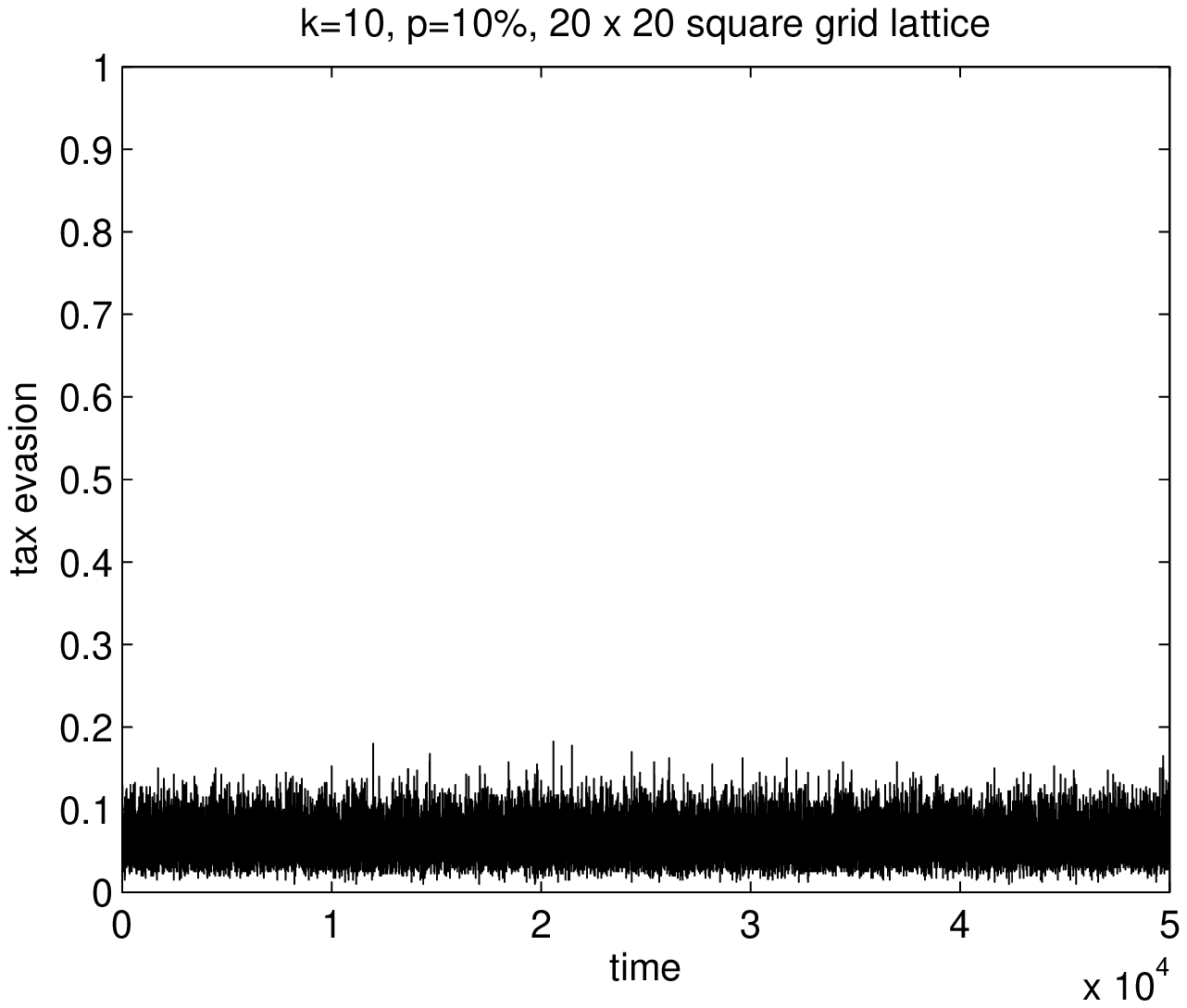}  
\hspace{-0.75cm}
\includegraphics[width=6.8cm]{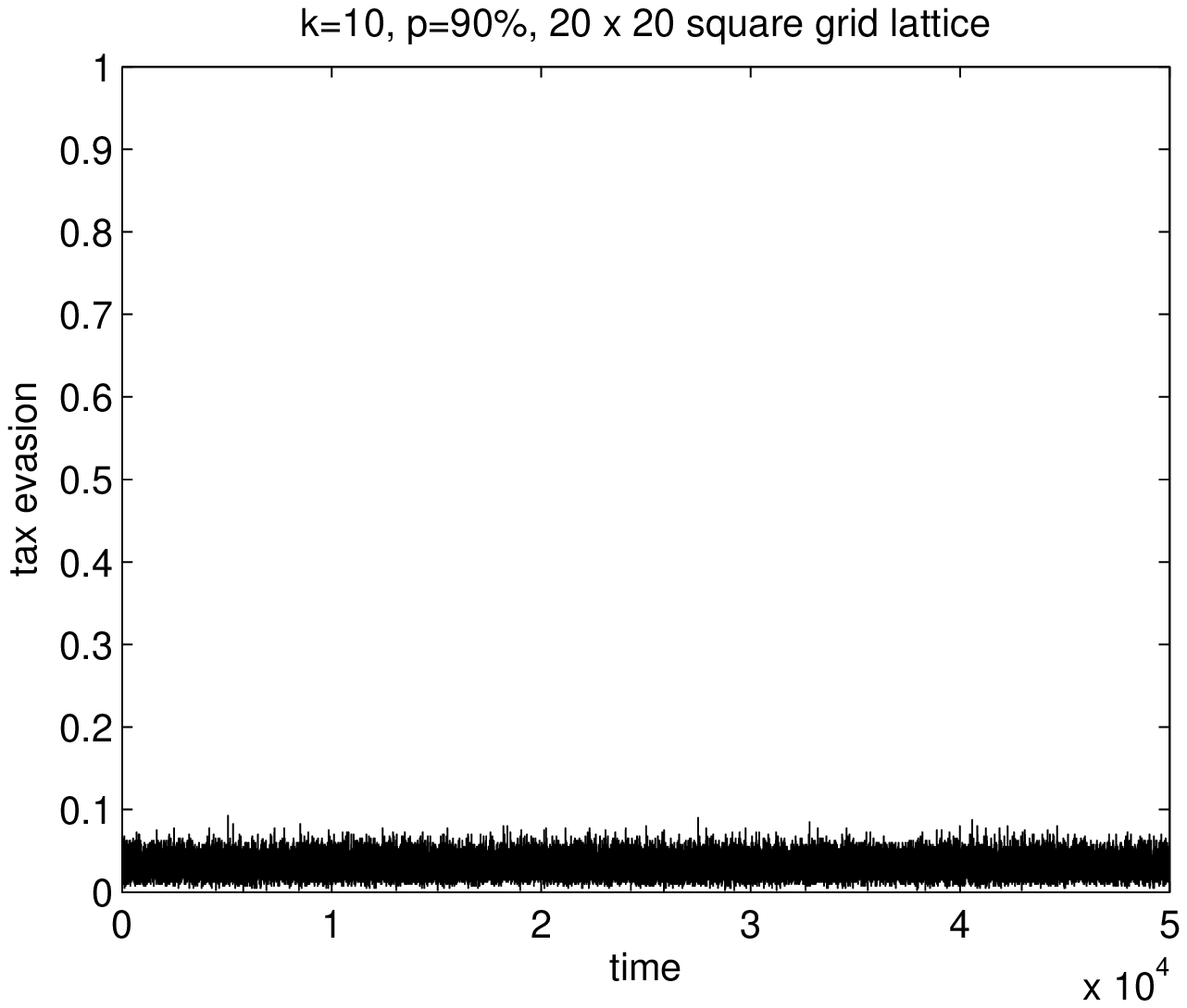} \\
\hspace{-0.8cm}
\hspace{-1.7cm}\includegraphics[width=6.8cm]{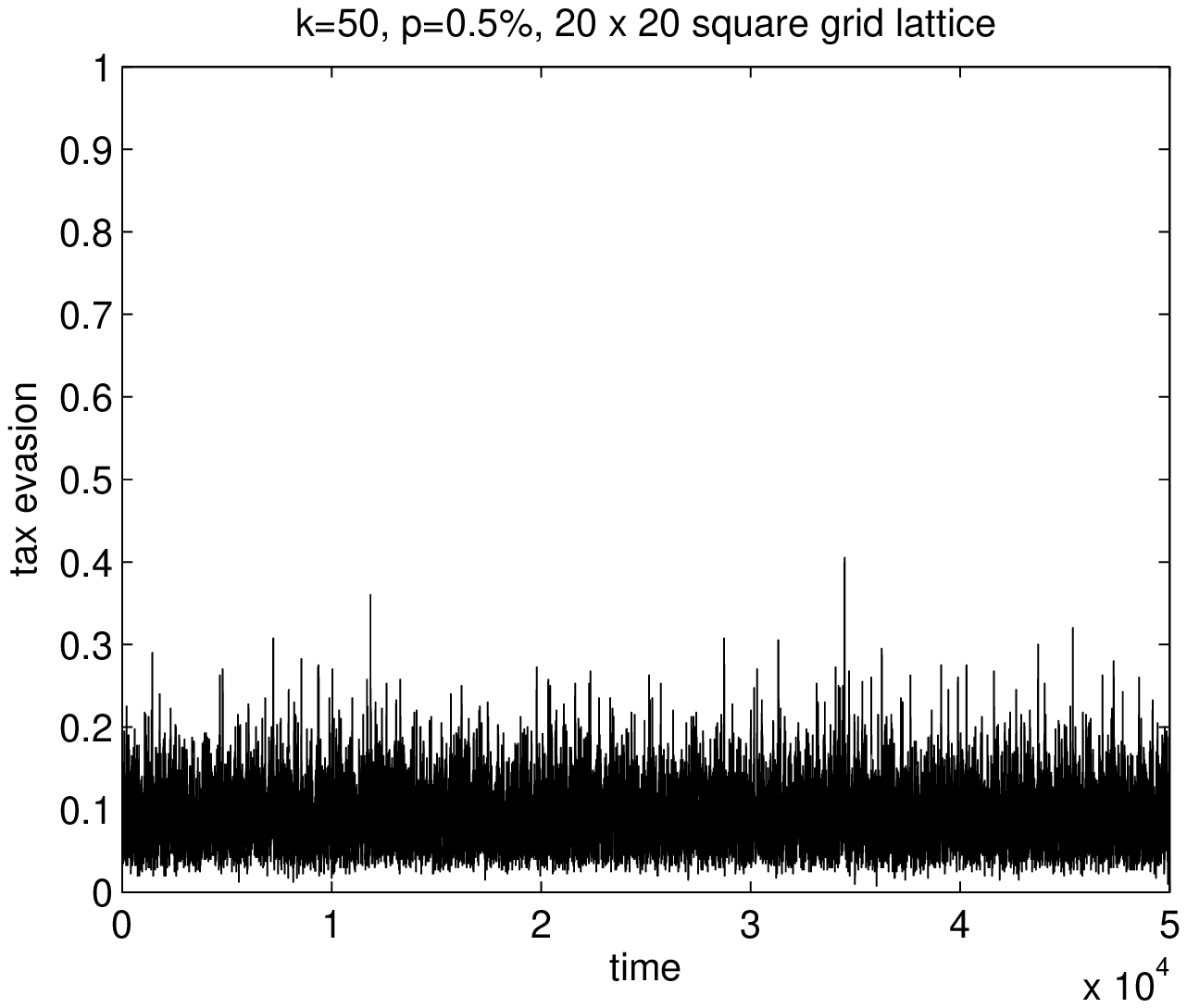}
\hspace{-0.75cm}
\includegraphics[width=6.8cm]{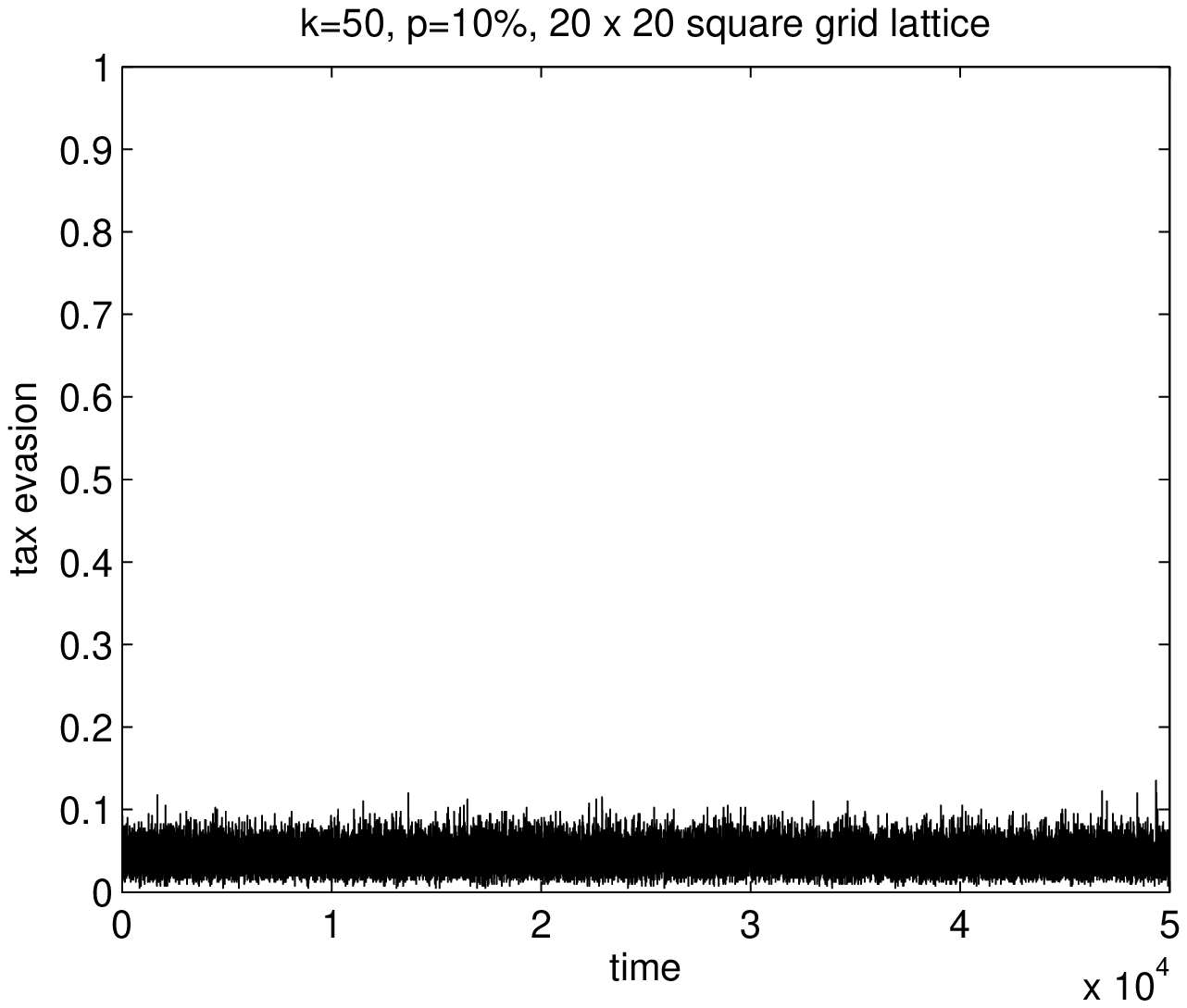}  
\hspace{-0.75cm}
\includegraphics[width=6.8cm]{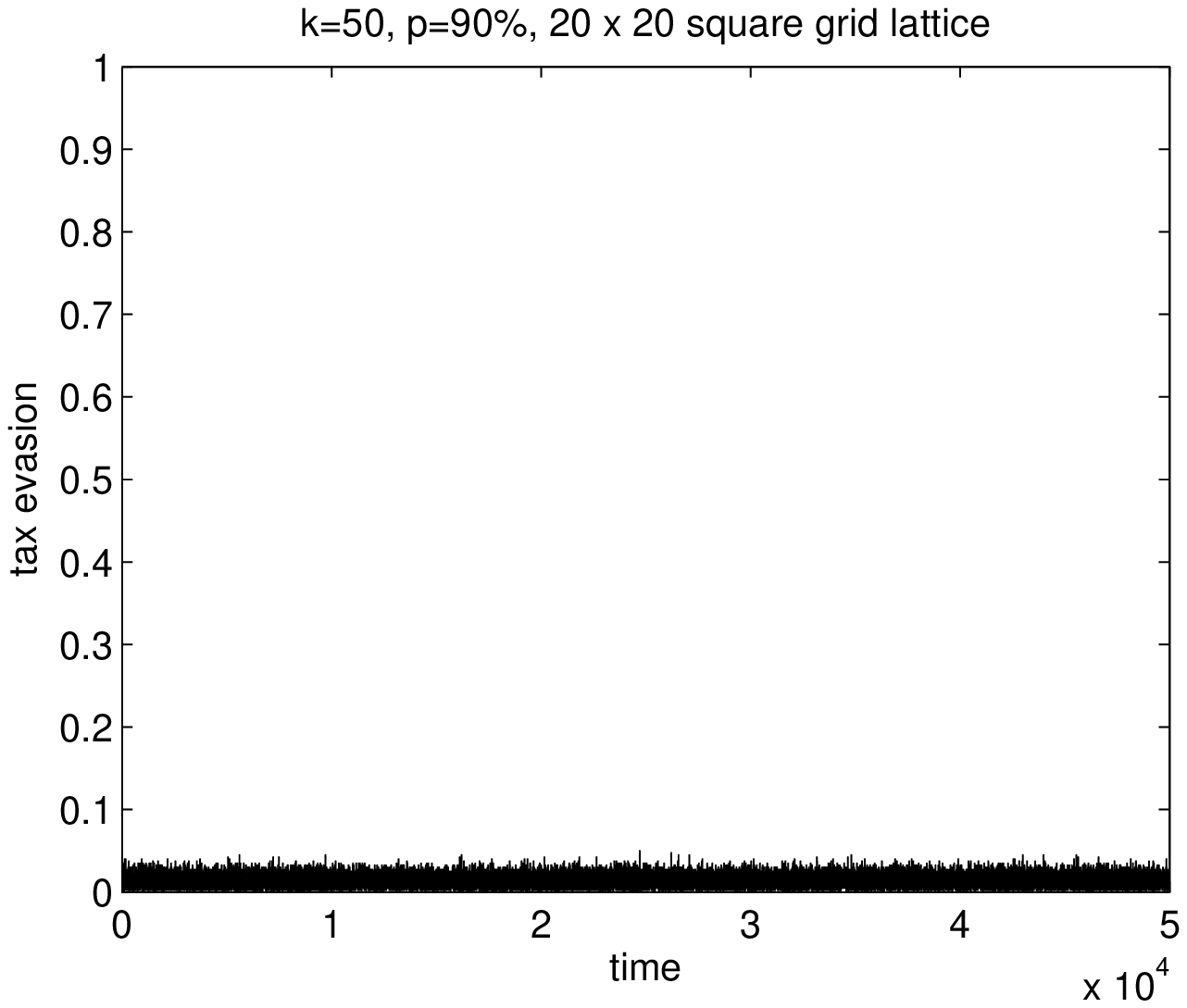} 
\end{tabular}
\end{center}
\vspace{0.1cm}
\begin{center}
{
{\noindent Figure 2: The square lattice model of tax evasion with various degrees of  enforcement. ($T=2.265$ and $50,000$ time steps)}}
\end{center}

\newpage
\vspace*{0.5cm}
\begin{center} 
\begin{tabular}{c}
\hspace{-0.8cm}							
\hspace{-1.7cm}\includegraphics[width=6.8cm]{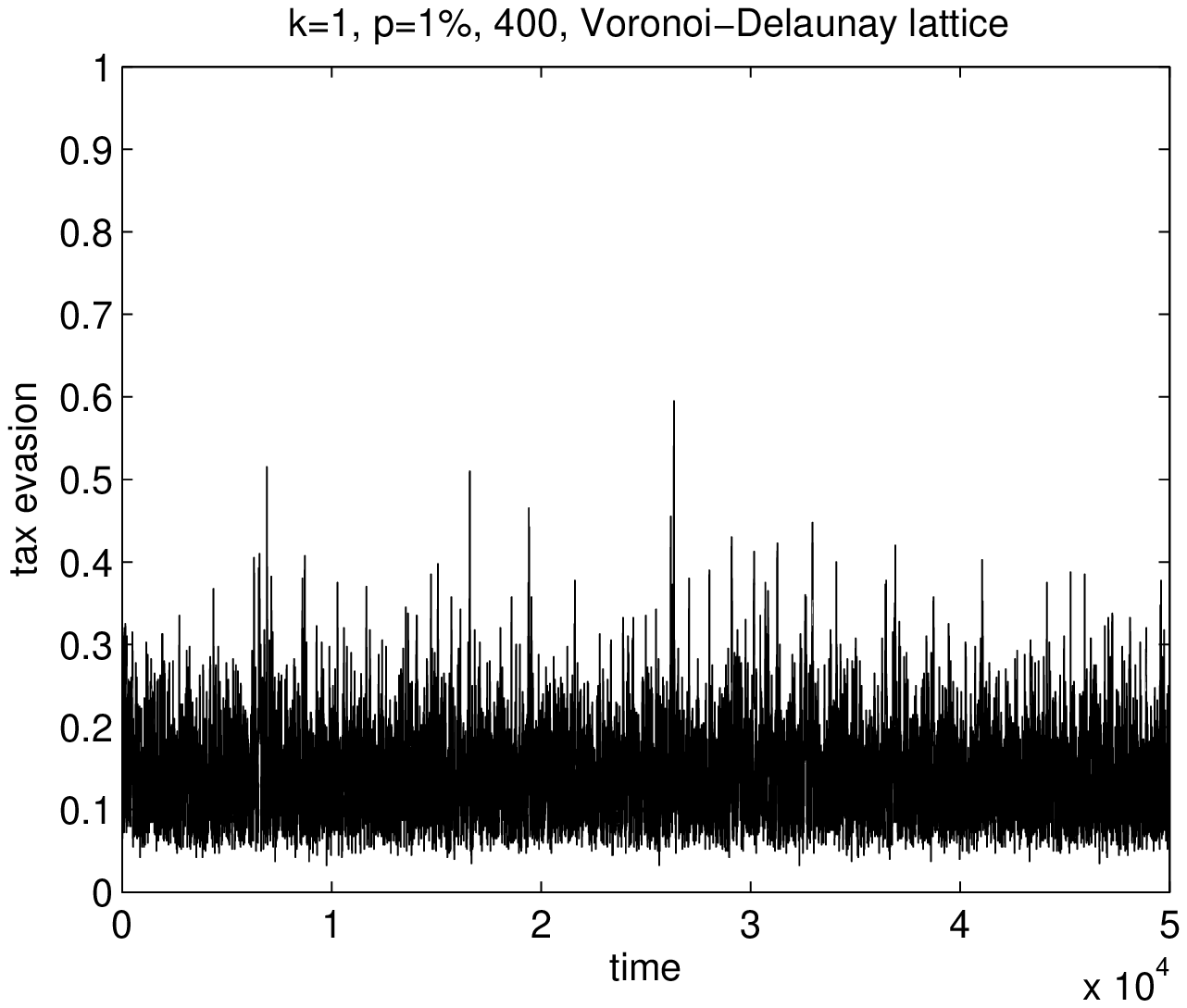}
\hspace{-0.75cm}
\includegraphics[width=6.8cm]{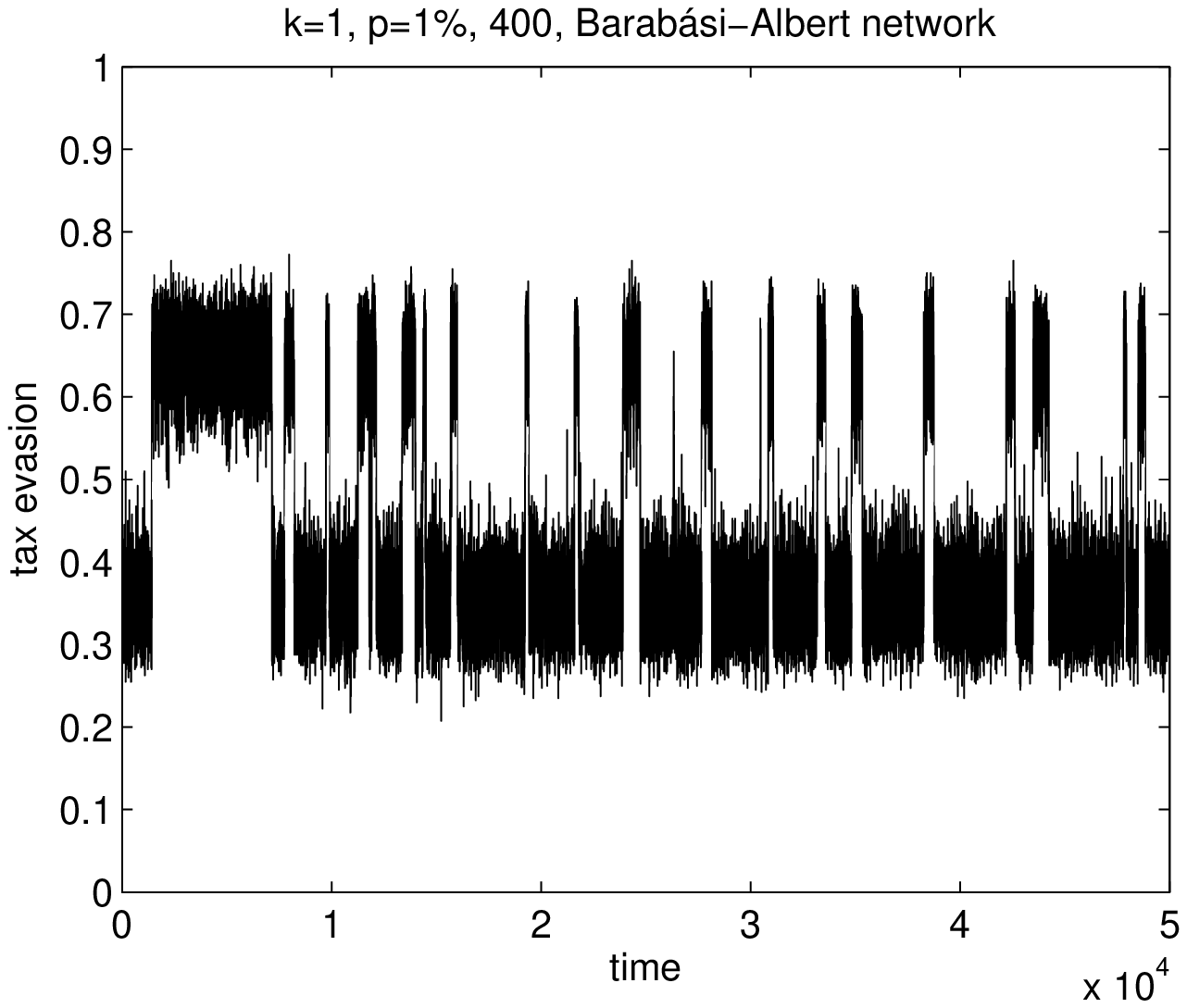}  
\hspace{-0.75cm}
\includegraphics[width=6.8cm]{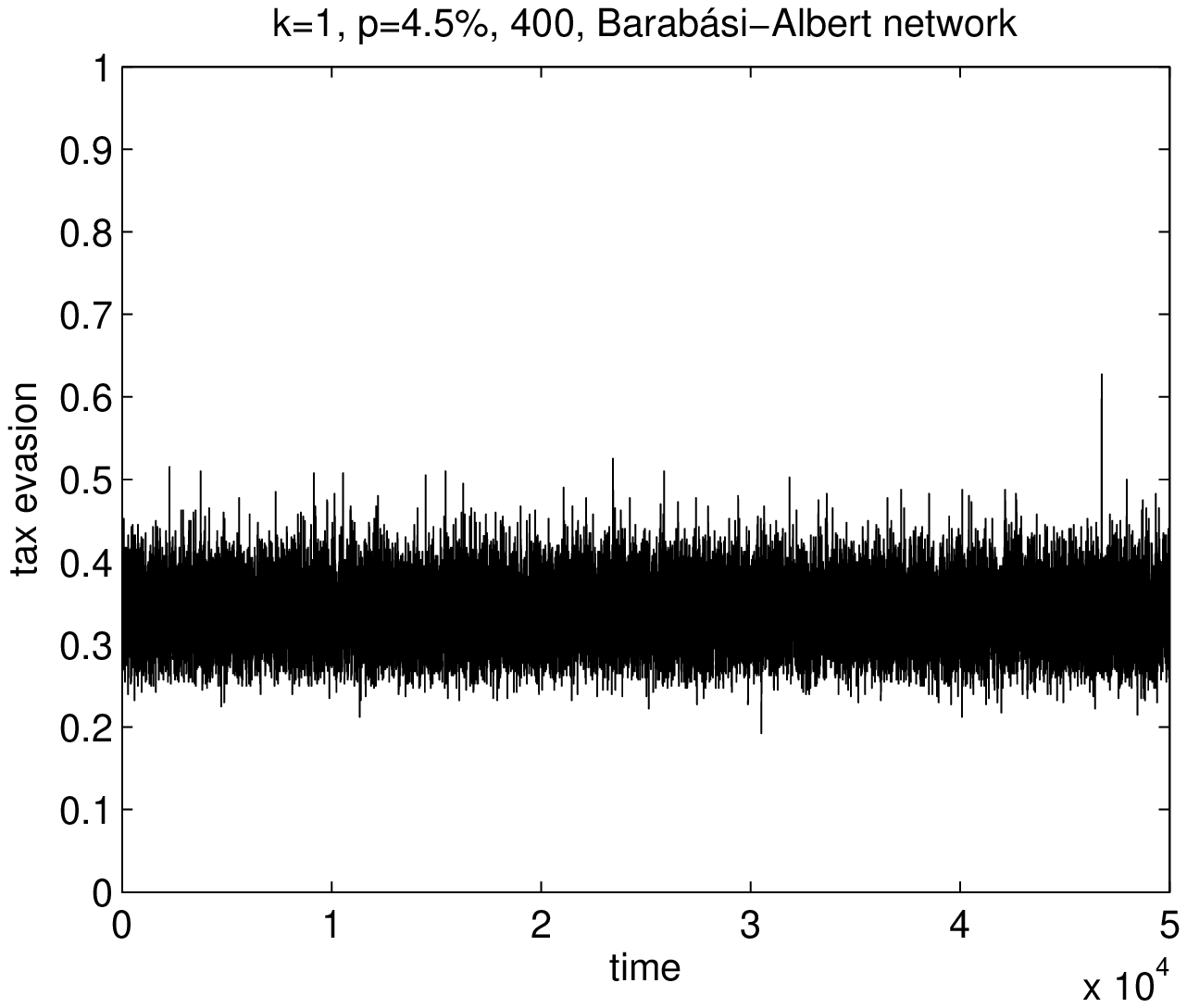} \\
\hspace{-0.8cm}									
\hspace{-1.7cm}\includegraphics[width=6.8cm]{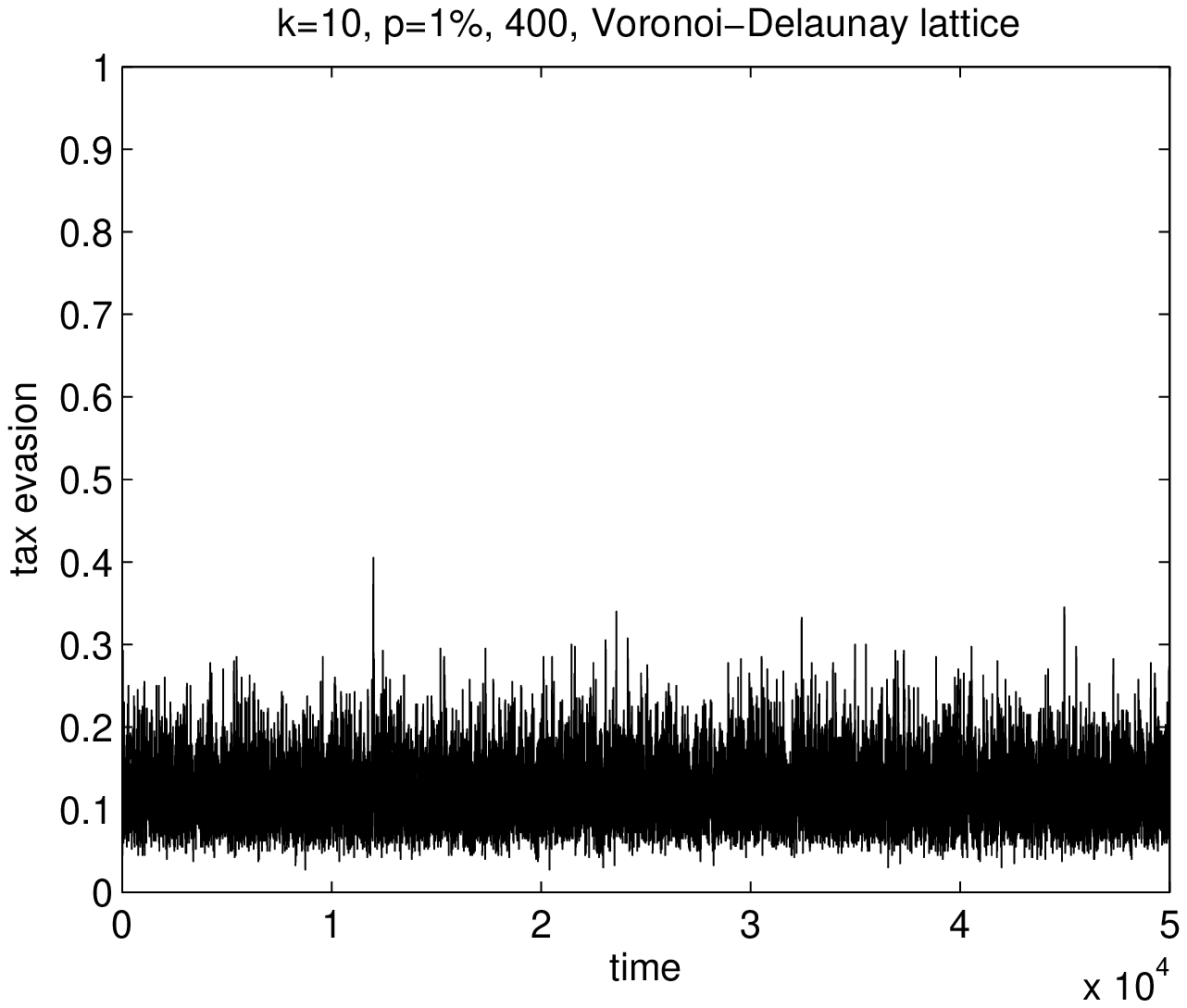}
\hspace{-0.75cm}
\includegraphics[width=6.8cm]{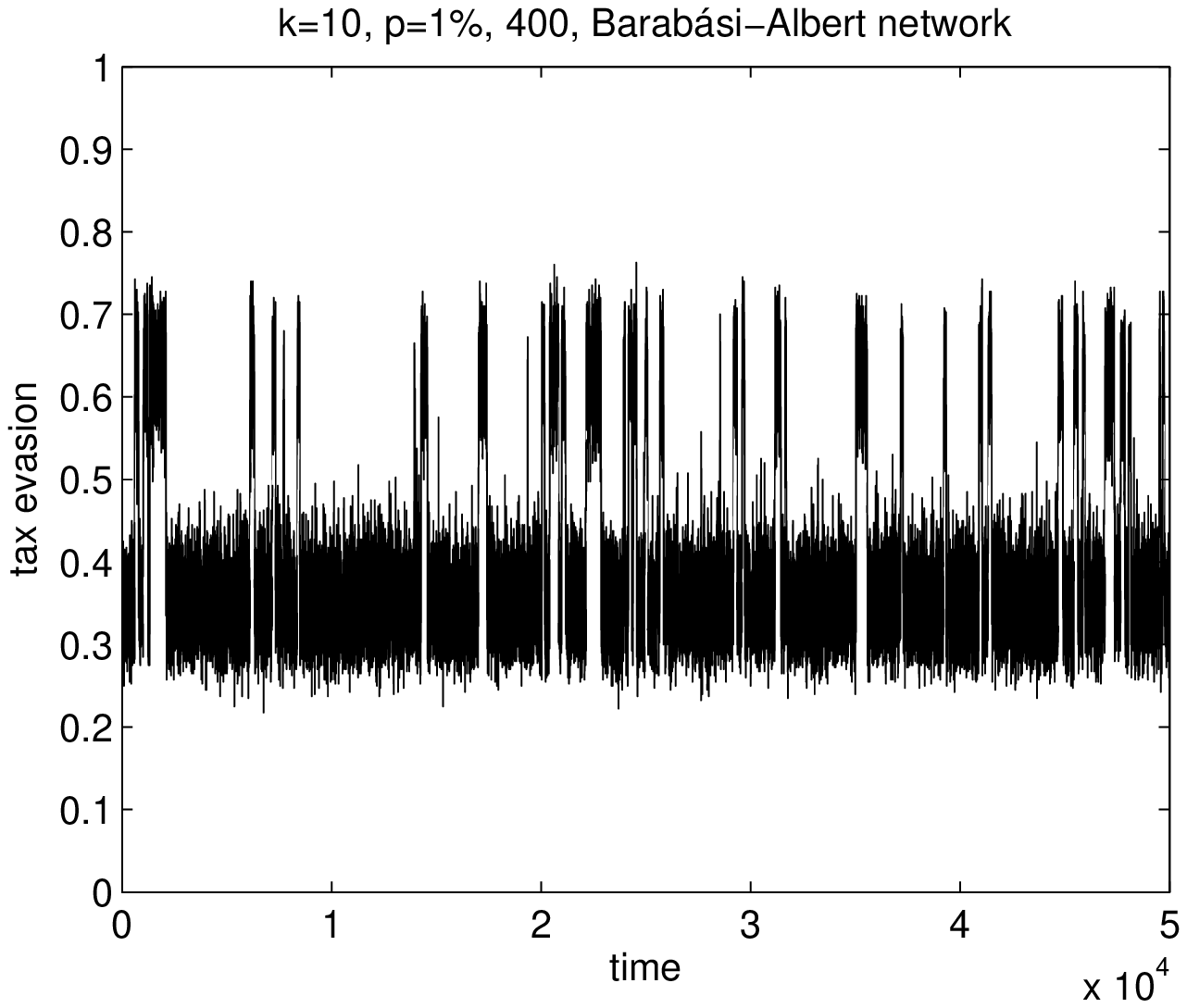}  
\hspace{-0.75cm}
\includegraphics[width=6.8cm]{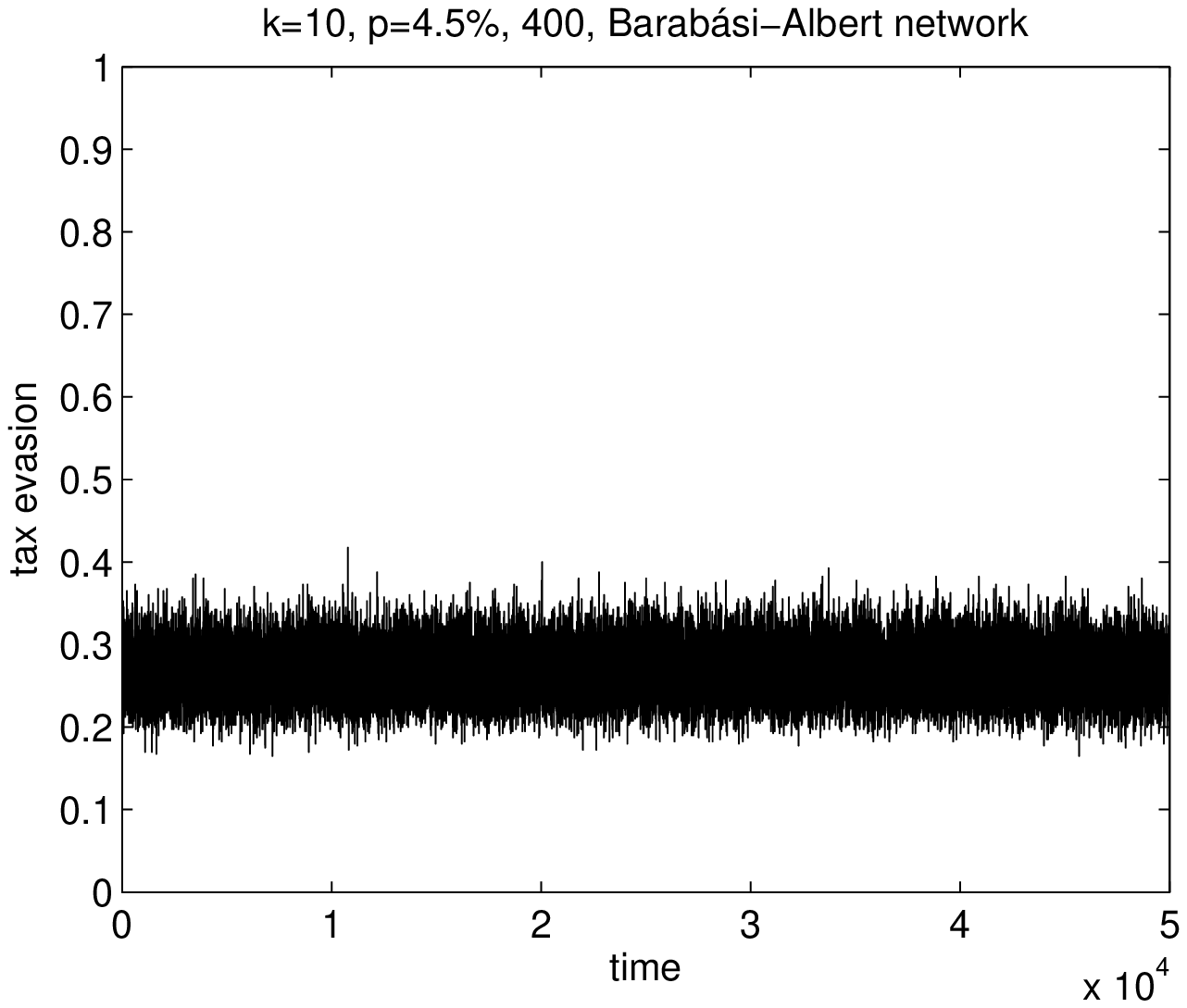} \\
\hspace{-0.8cm}										
\hspace{-1.7cm}\includegraphics[width=6.8cm]{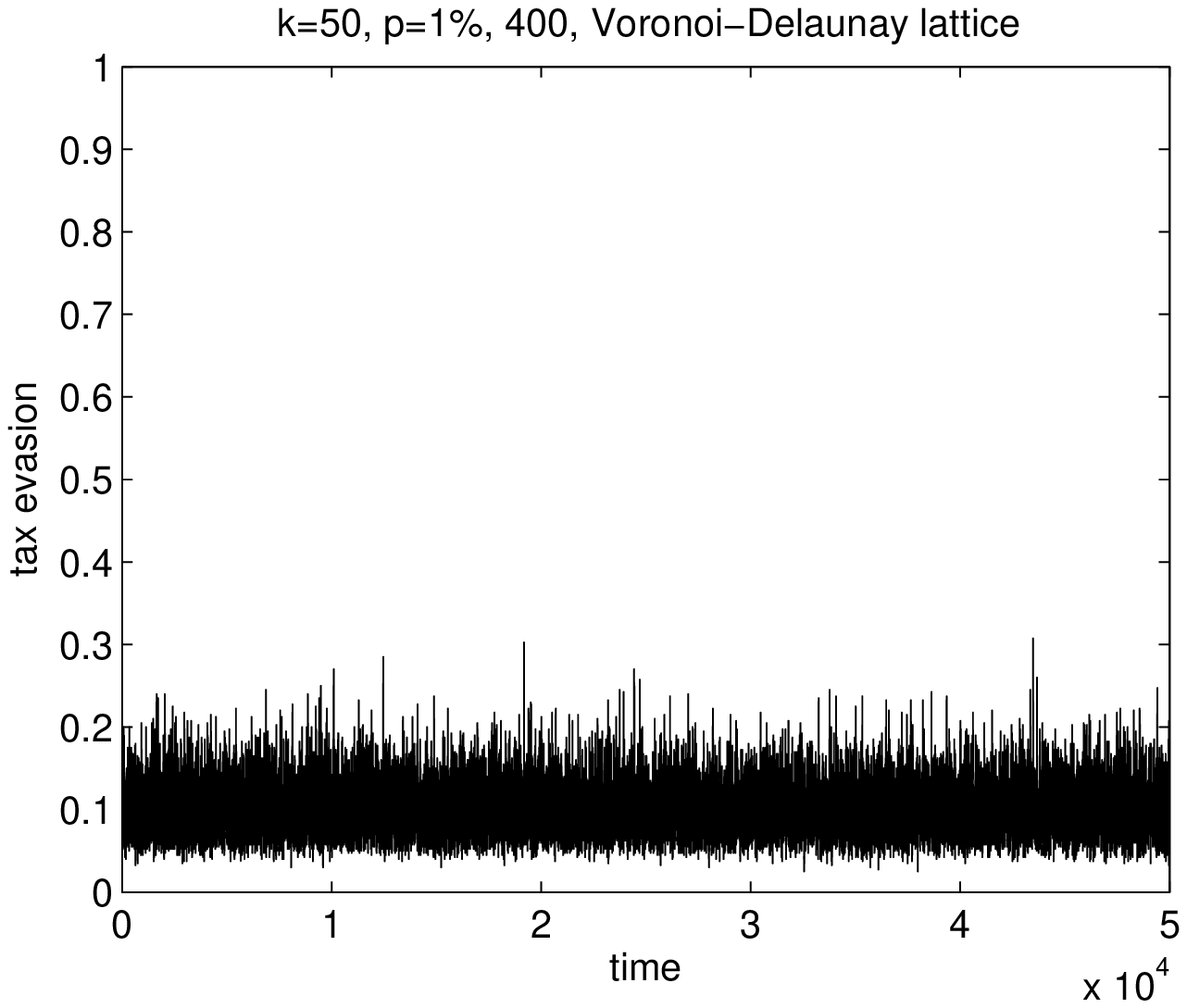}
\hspace{-0.75cm}
\includegraphics[width=6.8cm]{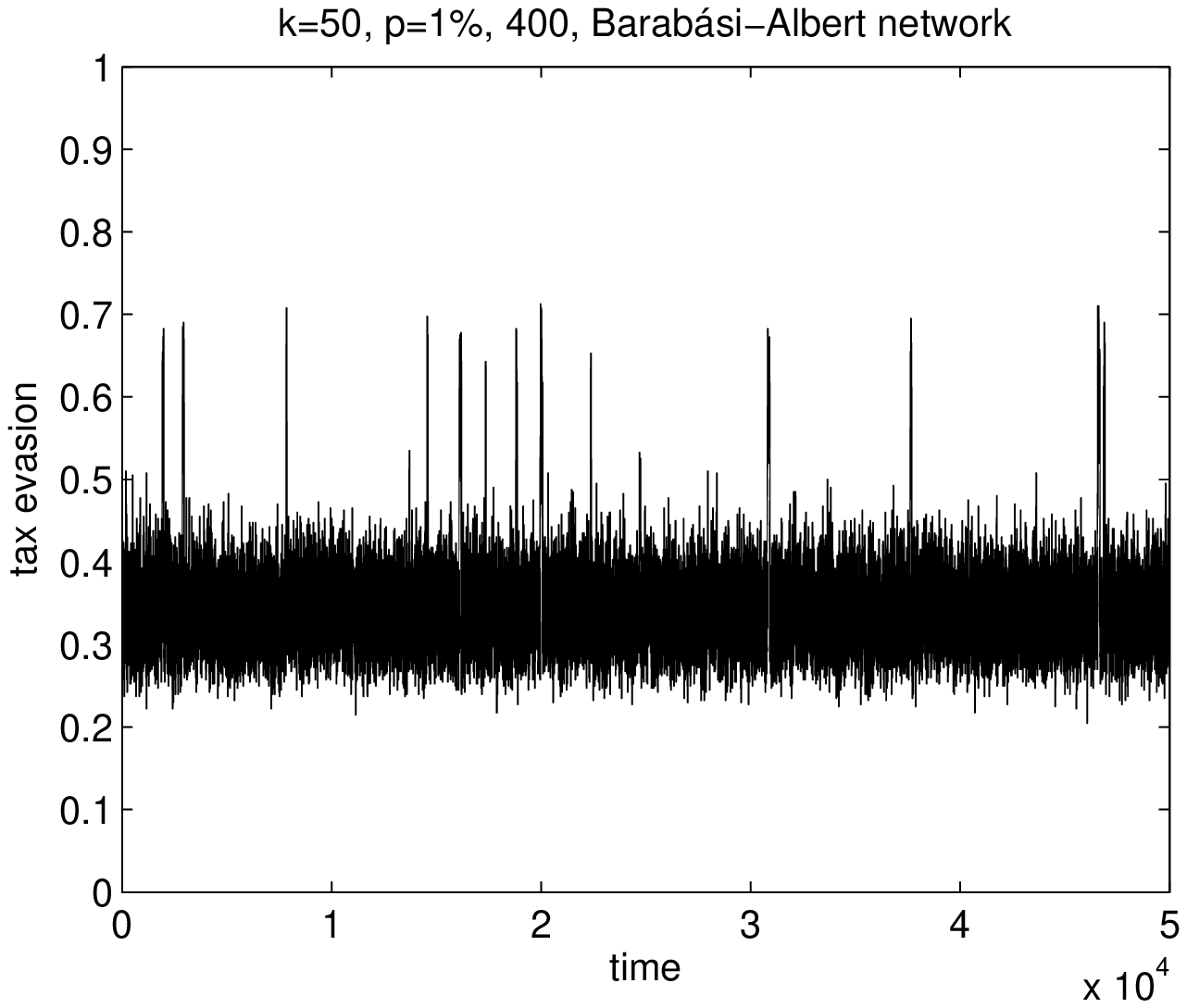}  
\hspace{-0.75cm}
\includegraphics[width=6.8cm]{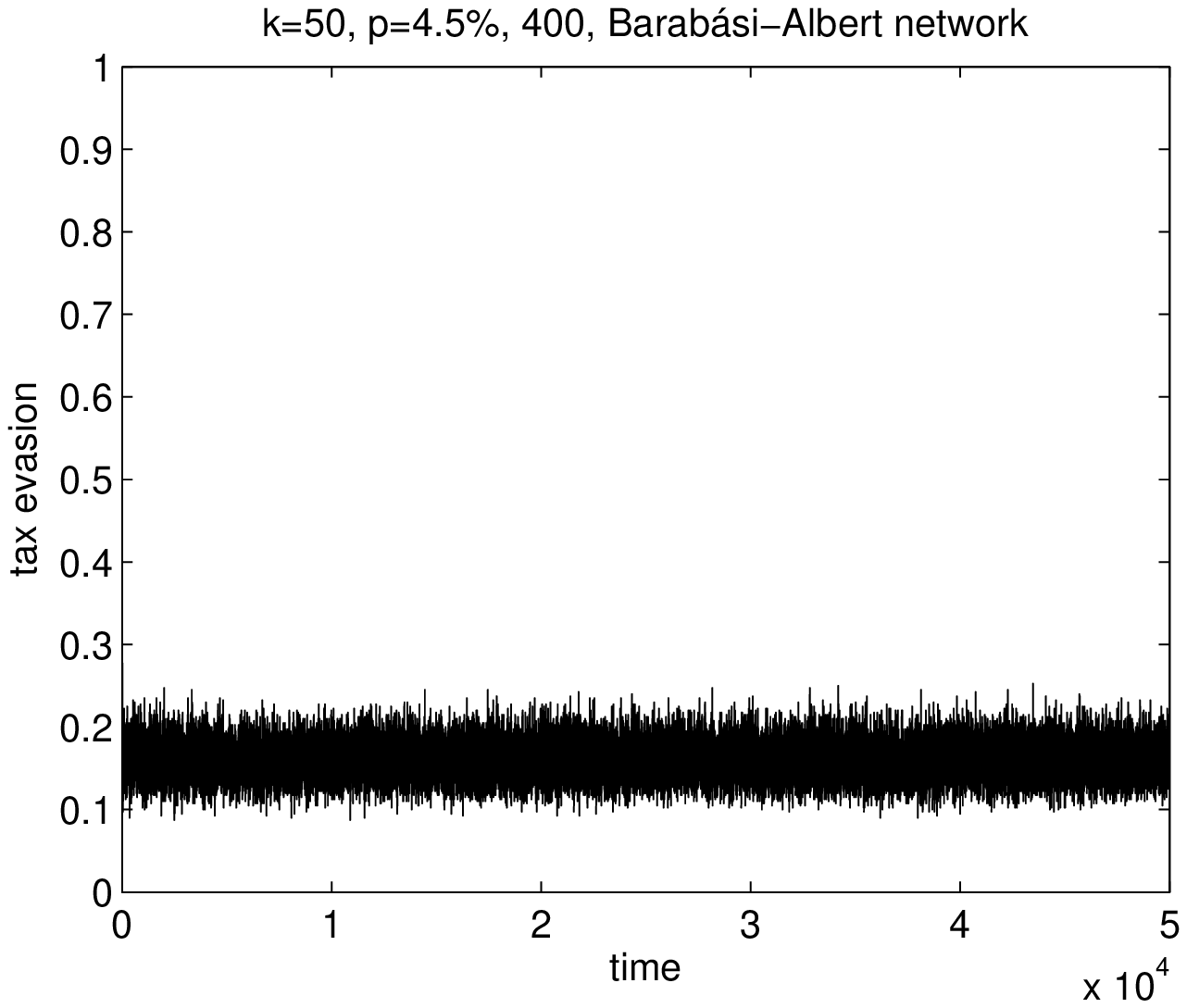} 
\end{tabular}
\end{center}
\vspace{0.5cm}

{\noindent Figure 3: The first column illustrates the resulting tax evasion dynamics for different enforcement regimes if the Voronoi-Delaunay network is used. The next two columns depict the tax evasion dynamics in the case of the Barab\'asi-Albert network.
Again, we use $50,000$ time steps.

\end{document}